\documentclass[preprint2,longabstract]{aastex}

\begin{document}
\title{A Theory about Electric Current and Heating in Plasma}
\author{Zhiliang Yang}
\affil{Department of Astronomy, Beijing Normal University, Beijing}
\email{zlyang@bnu.edu.cn}

\date{}


\label{firstpage}

\begin{abstract}
The traditional  generalized Ohm's law in MHD do not explicitly present the relation of electric currents and electric fields in fully ionized plasma, and lead to some unexpected concepts, such as ``the magnetic frozen-in plasma'', magnetic reconnection etc. In the single fluid
model, the action between electric current with magnetic field is not considered.
In the two-fluid model, the derivation is from the two dynamic equations of ions and electrons. However,
the electric current in traditional generalized Ohm's law depends on the velocities of the plasma, which could be decided by the two dynamic equations. The plasma velocity, eventually not free,  is inappropriately considered as free parameter in the traditional generalized Ohm's law.

In the present paper, we solve the balance equation can give exact solution of
the velocities of electrons and ions, and then derived the electric current in fully
ionized plasma. In the case ignoring boundary condition, there
is no electric current in the plane perpendicular to the magnetic field when external forces
are ignored.  The electric field in the plane perpendicular to magnetic field do not contribute to the electric currents, so do the induced electric field from the motion of the plasma across magnetic field. The lack of induced electric current will keep magnetic field in space unaffected. The velocity of the bulk velocity of the plasma perpendicular to magnetic field is not free, it is decided by electromagnetic field and the external forces.  We conclude that the bulk velocity of the fully ionized plasma is not coupled with the magnetic field. The motion of the plasma do not change the magnetic field in space, but the plasma will be confined by magnetic field.

Due to the confinement of magnetic field, the plasma kinetic energy will be transformed into plasma thermal energy by the Lamor motion and collisions between the same species of particles inside plasma. Because the electric field perpendicular to magnetic field do not contribute electric current, the variation of magnetic field will transfer energy directly into the plasma thermal energy by induced electric field. The heating of plasma could be from the kinetic energy and the variation of magnetic field.

\end{abstract}


\maketitle

\section{Introduction}

Recently Falthammar \citep{b01} commented on the unjustified use of the motion of magnetic field lines in plasma, following Alfven, who vigorously warned against the unjustified use of concept ``Frozen-in magnetic field" in his late years \citep{b02}. However, the scientists pay little attention to the suggestions. The concept of magnetic field line motion was applied extensively in space plasma physics and solar physics. Magnetic reconnection becomes  hot topic in explaining magnetic activities. The problem is  the generalized Ohm's law of MHD theory which shows the couple between velocity of plasma and the magnetic field.

 MHD(Magnetohydrodynamics) is the physical-mathematical framework that concerns the
dynamics of magnetic fields in electrically conducting fluids, e.g. in plasmas and liquid
metals. One of the most famous scholars associated with MHD was the Swedish physicist Hannes
Alfven, who receive the Nobel Prize in Physics in 1970 for fundamental work and discoveries
in MHD with fruitful applications in different parts of plasma physics. The notion of frozen-in
magnetic field is the result of his work in connection with the discovery of MHD waves \citep{b1}.

The central point of MHD theory is that conductive fluids can support the magnetic field. The
presence of magnetic fields leads to forces that in turn act on the fluid, typically a plasma,
thereby potentially altering the geometry and strength of the magnetic fields themselves.  Based
on the frame of MHD, a lot of theories come out, including MHD turbulence, MHD waves, magneto-convection, MHD reconnection, and MHD dynamo.

In fact, the MHD theory made little scientific progress in space science and astrophysics in the past decades. Problems are still in puzzle. The key problem of MHD theory comes from the generalized Ohm's law, which is believed to give the electric current in cosmic plasma.

\section{The traditional Generalized Ohm's Law and MHD equations}

In the standard non-relativistic form, the MHD equations consist of the basic conservation
laws of mass, momentum and energy together with the induction equation for the magnetic
field. The equations are, written in SI units:
\begin {equation}
\frac{\partial \rho}{\partial t}+\bigtriangledown \cdot \rho \bf{u}=0
\end {equation}
The equation of motion:
\begin {equation}
\frac{\partial (\rho \bf{u})}{\partial t}+\bigtriangledown \cdot (\rho \bf{u}\bf{u})=- \bigtriangledown p+\bf{j}\times \bf{B}+\bigtriangledown \cdot \sigma
\end {equation}
Where $\rho$ is the mass density and $\bf{u}$ the fluid bulk velocity, $p$ is the gas pressure, $\bf{B}$
the magnetic field, $\bf{j}$ the current density, and $\sigma$ is the viscous stress tensor.The equation for the internal energy, which is usually written as an equation for the
pressure $p$:
\begin {equation}
\frac{\partial p}{\partial t}+\bf{u}\cdot \bigtriangledown p +\gamma p\bigtriangledown \cdot \bf{u} =Q
\end {equation}
Where $Q$ comprises the effects of heating and cooling as well as thermal conduction and $\gamma$
is the adiabaticity coefficient. The above equation implies the equation of state of the
ideal ionized gas:
\begin {equation}
p=2(\rho /m_{i})k_{B}T
\end {equation}
It is satisfied for most dilute plasmas.

The induction equation, or Faraday's law:
\begin {equation}
\frac{\partial \bf{B}}{\partial t}=-\bigtriangledown \times \bf{E}=\bigtriangledown \times (\bf{u}\times \bf{B})+\eta \bigtriangledown ^2 \bf{B}
\end {equation}
Which is derived by inserting Ohm's law:
\begin {equation}
\bf{E}=-\bf{u}\times \bf{B} +\eta \bf{j}
\end {equation}
Here, $\eta$ is the electrical resistivity. In total, the MHD equation consist of two vector
and two scalar partial differential equation (or eight scalar equation) that are to be solved
simultaneously.

An early theoretical paper by \citet{b2} on magnetized plasma properties in the MHD
description contains in its physics sections criticisms about the applicability of ideal
or resistive MHD theory for plasmas. Specifically, ignoring the Hall
term in the generalized Ohm's law concerned, a simplification still made almost routinely in magnetic
fusion theory. The objections to classical MHD theory and their consequences give rise to the
development of Hall MHD and its application to laboratory and cosmic plasma \citep{b3}.

According to the MHD equation, we can find that the key point is the decision of the generalized
Ohm's law. \citet{b3} stressed that to retain the Hall term, using a two-fluid plasma description
is necessary. Then the generalized Ohm's law for fully ionized plasma can be derived.

For the macroscopic behavior of plasma, \citet{b4} gave the detailed discussion.
The basic equations are the equations of ions and electrons together with the Maxwell's equations,
the equation of continuity, and in the condition of $\frac{\partial}{\partial t}=0$, the generalized
Ohm's law is expressed as:
\begin {equation}
\bf{j}'=\sigma _{\parallel}\bf{E}_{\parallel}'+\sigma _{\perp}\bf{E}_{\perp}'+\sigma _{H} \bf{n}\times \bf{E}_{\perp}'
\end {equation}
Where,
\begin {equation}
\sigma_{\parallel}=\sigma=\frac{e^{2}n}{m_{e}\nu _{ei}}
\end {equation}

\begin {equation}
\sigma_{\perp}=\sigma \frac{1}{1+(\omega _{e} ^{2}/\nu_{ei}^{2})}
\end {equation}

\begin {equation}
\sigma_{H}=\sigma \frac{\omega _{e}/\nu _{ei}}{1+\omega _{e}^{2}/\nu _{ei} ^{2}}
\end {equation}

According to the above generalized Ohm's law, the current will depend on the conductivity of magnetized
plasmas. The magnetic field influence on the conductivity of the 'direct' current $\sigma _\perp$ and
that of Hall's current $\sigma _H$is determined by the parameter $\omega _{e}/v_{ei}$, which is nothing other
than the turning angle of an electron on the Larmor circle in the intercollisional time \citep{b5}.

In the case $\omega _{e}/\nu _{ei} \ll 1$, this corresponds to the weak magnetic field of dense cool
plasmas, so that the current is scarcely affected by the field:
\begin {equation}
\sigma_{\perp} \approx \sigma_{\parallel} \approx \sigma, \frac{\sigma_{H}}{\sigma} \approx \omega _{e}/\nu _{ei} \ll 1
\end {equation}

Thus in a frame of reference associated with the plasma, the usual Ohm's law with isotropic conductivity
holds.

In the opposite case, when the electrons spiral freely between collisions, $\omega _{e}/\nu _{ei}\gg 1$, we will get the strong magnetic field and hot rarefied plasma. This plasma is termed the magnetized one. It is
frequently encountered under astrophysical conditions. In this case,
\begin {equation}
\sigma_{\parallel} \approx (\omega _{e}/\nu _{ei})\sigma _{H} \approx (\omega _{e}/\nu _{ei})^{2}\sigma _{\perp}
\end {equation}

Hence in the magnetized plasma, for example in the solar corona, $\omega _{\parallel} \gg \omega _{
H} \gg \omega _{\perp}$. The impact of the magnetic field on the direct current is especially strong
for the component resulting from the electric field $E'_{\perp}$. The current in the $E'_{\perp}$
direction is considerably weaker than it would be in the absence of a magnetic field.

In the cosmic conditions, the generalized Ohm's law assumes the form of Eq.(5) can be used as the
approximation of Eq.(7) \citep{b5,b6}. This seems to be perfect for the consistence of single fluid and two-fluid plasma model.

However, \citet{b4} pointed out the roles of the generalized Ohm's law and the dynamic equations have reversed roles from the usual custom, the equation of motion determines the current density, while the generalized Ohm's law determines the velocity. But in the research, the generalized Ohm's law is applied independently from the motion equation.

The generalized Ohm's law with the form of Eq.(6) and Eq.(7) was accepted and widely applied in
space science and astrophysics. According to the generalized Ohm's law with the form of Eq.(6)
and the Farady's law, we can get the magnetic induction equation:
\begin {equation}
\frac{\partial \bf{B}}{\partial t}
=\bigtriangledown \times (\bf{u} \times \bf{B})+ \eta \bigtriangledown ^{2} \bf{B}
\end {equation}
where $\eta =\frac{1}{4 \pi \omega}$ is the magnetic diffusivity. When the magnetic diffusivity
$\eta$ is infinite small, the plasma is called ideal MHD. Alfven noted that the motion of matter may
couple to the deformation of the magnetic field such that the field lines follow the motion of matter,
and devoted this "frozen-in magnetic field lines". The "frozen-in magnetic field lines" theorem and
its corollary "ideal MHD" are widely used in space plasma physics.

One of the most important deviation from ideal MHD is magnetic reconnection, which is the merging
of magnetic field lines, as invented by \citet{b7} and \citet{b8}, and later applied to the
Earth's magnetosphere by \citet{b9}. However, the physics involved in the magnetic field line merging
remains poorly understood.

Though the ideal MHD can be used to simulate the energy transfer from magnetic field to plasma kinetic
energy \citep{b10}, the existence of $E_{\parallel}$ along auroral magnetic field lines,
upward directed, as well as downward directed, is recognized \citep{b11}. It requiring a
finite parallel conductivity along the magnetic field $\bf{B}$. This may lead to the breakdown of
the single fluid concept of MHD \citep{b12}.

The derivation of the generalized Ohm's law can be found in many classic books and literatures about
plasma. The derivations include the discussions on single fluid plasma (e.g.\citet{b13}), the two-fluid (electrons and ions) \citep{b5,b4} plasma and the three-fluid
(electrons,ions and neutral particles).

However, through  carefully examination of the derivation, we find the electric current in previous generalized Ohm's law
is not explicitly presented. The velocity of the plasma should be decided by the electromagnetic field and external forces, it is not a free parameter.

\section{The electric current in fully ionized plasma}

To get the generalized Ohm's law in plasma, we consider the plasma as fluids, which means the distribution
of the particle velocity is Maxwellian distribution. The generalized Ohm's law is under the condition that
$\frac{\partial j}{\partial t}=0$ and the macroscopic velocities of the fluid (or electron fluid and ion
fluid) $\frac{\partial u}{\partial t}=0 $ in the plasma \citep{b4}.

The generalized Ohm's law for single fluid is derived simply from the reference system transformation \citep{b13}. It is first criticized by \citet{b2} for the lacking of the Hall term. The mistake
eventually is from the physical consideration. In the single fluid case, the magnetic field in the reference
system is not considered for the reason that the fluid is still. However, the impact of the magnetic
field is especially strong for the components resulting from the electric field $E'_{\perp}$ perpendicular
to the magnetic field in the fluid system. The primary effect of the electric field $E'_{\perp}$ in
the presence of the magnetic field is not the current in the direction $E'_{\perp}$, but rather
the electric drift in the direction perpendicular to both magnetic field and $E'_{\perp}$.

A fully treatment of the plasma can be started from the kinetic equation for each species of the
plasma. For the fully ionized plasma, they are the kinetic equations for electrons and ions. One
then integrates these equations over the phase space, defines macroscopic quantities, and derives
various moment equation for each species \citep{b14,b15,b16}.
These moment equations and macroscopic quantities describe each species as a fluid without invoking
the motion of each individual particle. These equations can include the interaction among different
species. They are, for ions,
\begin {equation}
n_{i}m_{i}\frac{D\bf{u}_{i}}{Dt}=-\bigtriangledown P_{i}+en_{i}(\bf{E}+\bf{u}_{i}\times \bf{B})+\bf{F}_{i}-
n_{e}m_{e}\nu _{ie}(\bf{u}_{i}-\bf{u}_{e})
\end {equation}
and for electrons,
\begin {equation}
n_{e}m_{e}\frac{D\bf{u}_{e}}{Dt}
=-\bigtriangledown P_{e}-en_{e}(\bf{E}+\bf{u}_{e}\times \bf{B})+\bf{F}_{e}
-n_{i}m_{e}\nu _{ei}(\bf{u}_{e}-\bf{u}_{i})
\end {equation}
Where the parameters $n_{i},n_{e},P_{i},P_{e},u_{i},u_{e},\nu _{ie}=\nu _{ei}$ are the density of ions,
density of electrons, pressure of the ion fluid, pressure of the electron fluid, macroscopic velocity of the ion fluid, macroscopic velocity of electron fluid, collision frequency of ions and electrons. And vector $F_{i}$, and $F_{e}$ denotes other forces. such as gravity, centrifugal force, and Coriolis force, exerting on the ion fluid and the electron fluid. We assume that the ions are singly charged and that charges neutrality holds in the plasma $n_{i}=n_{e}=n$, and ignoring possible wave-particle collisions.

To get the Ohm's law, we neglect the effects from time derivatives, pressure gradient forces, and
additional forces, which are referred as "other forces".

In the derivation of generalized Ohm's law, \citet{b5} considered the time derivatives and additional
forces as part of the effective electric field. When there are magnetic fields, the effect of
external forces acting on charged fluid is completely different to the electric field. The effects
can be seen from the drifts of charged particles in the magnetic field. In plasma, the drift due to
the electric field causes a macroscopic drift velocity, but there is no current. However,  the drift
due to external forces will cause electric current in plasma - the drift current \citep{b17}. In
this paper, we will first neglect the effect of other forces. The discussion with other forces is in the
later part.

From equations (14) and (15), Ignoring the external forces, the two fluid force balance equations for ions and electrons in steady state in our discussion will be:
\begin {equation}
en(\bf{E}+\bf{u}_{i} \times \bf{B})=nm_{e}\nu _{ei}(\bf{u}_{i}-\bf{u}_{e})
\end {equation}

\begin {equation}
-en(\bf{E}+\bf{u}_{e} \times \bf{B})=nm_{e}\nu _{ei}(\bf{u}_{e}-\bf{u}_{i})
\end {equation}
The two equations are coupled by the electron-ion collision terms. We define the electric current density as,
\begin {equation}
\bf{j}=en(\bf{u}_{i}-\bf{u}_{e})
\end {equation}
In all the previous derivations, the authors define the macroscopic velocity of plasma flow as\citep{b4,b5,b15}.
\begin {equation}
\bf{u}=\frac{m_{i}\bf{u}_{i}+m_{e}\bf{u}_{e}}{m_{i}+m_{e}}
\end {equation}
With the relationship of $m_{i} \gg m_{e}$, the plasma flow velocity becomes
\begin {equation}
\bf{u}=\bf{u}_{i}+\frac{m_{e}}{m_{i}}\bf{u}_{e}
\end {equation}
By substituting (18) and (20) into equations (16) and (17), then subtracting (16) from (17), one
then gets the form of Ohm's law the same as equation(7).

In fact, the Ohm's law is not completed. Besides the approximation of the equation (20), we can
have the additional relation for the current. Adding equations (16) and (17) together directly will give the result for $\bf{j}\times \bf{B}=0$. Some authors have discussed the results. Early in 1956, Spitzer discussed the dynamics of the full ionized plasma, and a later version in 1962 \citep{b4}. In the discussion of the macroscopic behavior of a plasma, Spitzer obtains the simpler equations in the conditions that ignoring the terms in $m_{e}/m_{i}$,$\frac{\partial \bf{j}}{\partial t}$ and
$\frac{\partial \bf{u}}{\partial t}$, considering the changes so slow that inertial effects are negligible.
With the gravity and the pressure gradient kept, Spitzer obtained the relationship of,
\begin {equation}
\bigtriangledown p=\bf{j} \times \bf{B}-\rho\bigtriangledown \phi
\end {equation}
Where p is the pressure,$\rho$ is the density and $\phi$ is the gravity potential. Spitzer suggested
that the equation (21) be the equation of motion and determine the current density \citep{b4}.

However, in the case that we derive the generalized Ohm's law, the pressure gradient and gravity
potential is always ignored. This will lead to the result of,
\begin {equation}
\bf{j}\times \bf{B}=0
\end {equation}
In this case, there will be no any current perpendicular to the magnetic field, a contradicted
result to the generalized Ohm's law of equation (7). If the pressure gradient is considered, the
generalized Ohm's law will be different, which will be discussed later in Part 3.

In principle, we don't need to get the solution by above method. The balance equations (16) and (17)
can completely determine the velocities of ion fluid and electron fluid in the slowly varied
electromagnetic field system.

Setting the magnetic field in the direction of $z$, so $\bf{B}=B_z$, we can have the electric field
perpendicular to magnetic field in the direction $x$, that is $\bf{E}_{\perp}=\bf{E}_{x}$, and $\bf{E}_{\parallel}=
\bf{E}_{z}$. The direction $-\bf{E}\times \bf{B}$ is the direction of $y$. Then we can have equation (16)
and (17) written in the form of components in the direction of $x$, $y$ and $z$.

\begin{equation}
\frac{eE_x}{m_e}+\omega_eu_{iy}-\nu_{ei}(u_{ix}-u_{ex})=0
\end{equation}

\begin{equation}
-\omega_eu_{ix}-\nu_{ei}(u_{iy}-u_{ey})=0
\end{equation}

 \begin{equation}
-\frac{eE_x}{m_e}-\omega_eu_{ey}-\nu_{ei}(u_{ex}-u_{ix})=0
 \end{equation}

\begin{equation}
\omega_{e}u_{ex}-\nu_{ei}(u_{ey}-u_{iy})=0
\end{equation}

\begin{equation}
\frac{e}{m_e}(E_z)-\nu _{ei}(u_{iz}-u_{ez})=0
\end{equation}

\begin{equation}
-\frac{e}{m_e}(E_z)-\nu _{ei}(u_{ez}-u_{iz})=0
\end{equation}

Where $\omega _{i}=\frac{eB}{m_{i}c},\omega _{e}=\frac{eB}{m_{e}c}$. The solutions of equation (23), (24), (25), (26) are,
\begin{equation}
u_{ex}=u_{ix}=0
\end{equation}
And
\begin{equation}
u_{ey}=u_{iy}=\frac{\bf{E}\times \bf{B}}{\bf{B}^{2}}
\end{equation}

As we have defined, the current in the plane perpendicular to the magnetic field is zero, since the ions and electrons have exactly the same drift velocity . It is consistent with
the equation of (22) as the other forces ignored (the force-free case). Then we have only the current in the direction parallel to the magnetic field. Along the magnetic
field line, it is the normal Ohm's law since there is not effect by the magnetic field. So we can
conclude that in the fully ionized plasma, when the other forces besides the electromagnetic field
are ignored, the electric current in the plasma will be:
$$
\bf{j}=\sigma \bf{E} _{\parallel}=\sigma \frac {\bf{B}(\bf{E}\cdot \bf{B})}{B^2}
$$
To be consistent with the common Ohm's law, we notice the relation:

$$
\frac {\bf{B}(\bf{E}\cdot \bf{B})}{B^2} = \bf{E} + \frac {(\bf{E}\times \bf{B})\times \bf{B}}{B^2}
$$

Then we can have the generalized Ohm's law as:
\begin{equation}
\bf{j} = \sigma [\bf{E} + \frac {(\bf{E}\times \bf{B})\times \bf{B}}{B^2}]
\end{equation}
and the plasma velocities is determined by the following relation,
\begin{equation}
\bf{v} = \frac {m_e\bf{v}_e + m_i\bf{v}_i}{m_e + m_i} = \frac{\bf{E}\times \bf{B}}{\bf{B}^{2}}
\end{equation}

In the above method, we can decide the electric current and the plasma velocity by the electric field and the magnetic field in any reference system.

\section{The electric current from external forces in fully ionized plasma}

    The previous Ohm's law is always confused when the external forces included. In this part, we
will discuss the case when external forces are included in the fully ionized plasma. We consider
the pressure gradient, gravity or friction as the external forces $f_{i}$ and $f_{e}$ for the fluid
of ion and electron respectively.
$
\bf{f}_{i}=\frac{\bf{F}_{i}-\bigtriangledown P_{i}}{nm_{e}}
$, $
\bf{f}_{e}=\frac{\bf{F}_{e}-\bigtriangledown P_{e}}{nm_{e}}
$. In the nearly steady case, the balance equations for the ion fluid and electron fluid in x, y plane will be:
\begin {equation}
\frac{e}{m_{e}}(\bf{E}+\bf{u}_{i} \times \bf{B})+\bf{f}_{i}=\nu _{ei}(\bf{u}_{i}-\bf{u}_{e})
\end {equation}
\begin {equation}
-\frac{e}{m_{e}}(\bf{E}+\bf{u}_{e} \times \bf{B})+\bf{f}_{e}=\nu _{ei}(\bf{u}_{e}-\bf{u}_{i})
\end {equation}
In the direction of magnetic field will be:
\begin{equation}
\frac{e}{m_{e}}(E_{z})+{f}_{iz}=\nu _{ei}(u_{iz}-u_{ez})
\end{equation}
\begin{equation}
-\frac{e}{m_{e}}(E_{z})+{f}_{ez}=\nu _{ei}(u_{ez}-u_{iz})
\end{equation}

Here, we assume the magnetic field in the direction of $z$, the electric field perpendicular to
magnetic field in the direction $x$, and $-\bf{E}\times\bf{B}$ is in the direction of $y$. And equations (32), (33), (34) and (35) can be written as:
\begin{equation}
\frac{eE_{x}}{m_{e}}+\omega_{e}u_{iy}-\nu_{ei}(u_{ix}-u_{ex})+f_{ix}=0
\end{equation}
\begin{equation}
-\omega_{e}u_{ix}-\nu_{ei}(u_{iy}-u_{ey})+f_{iy}=0
\end{equation}
\begin{equation}
-\frac{eE_{x}}{m_{e}}-\omega_{e}u_{ey}-\nu_{ei}(u_{ex}-u_{ix})+f_{ex}=0
\end{equation}
\begin{equation}
\omega_{e}u_{ex}-\nu_{ei}(u_{ey}-u_{iy})+f_{ey}=0
\end{equation}
\begin{equation}
\frac{e}{m_{e}}(E_{z})-\nu _{ei}(u_{iz}-u_{ez})+f_{iz}=0
\end{equation}
\begin{equation}
-\frac{e}{m_{e}}(E_{z})-\nu _{ei}(u_{ez}-u_{iz})+f_{ez}=0
\end{equation}

Then, from equations (37), (38), (39) and (40)  we can get the velocities of ions and electrons in the plane perpendicular to magnetic field as,
\begin{equation}
u_{ex}=\frac{(f_{ey} + f_{iy})\nu_{ei} - f_{ex}\omega_e}{\omega _e ^2}
\end{equation}

\begin{equation}
u_{ey}= \frac{(f_{ex} + f_{ix})\nu_{ei} + f_{ex}\omega _e}{\omega _e^2} - v_s
\end{equation}

\begin{equation}
u_{ix}=\frac{(f_{ex}+f_{ix})\nu_{ei} + f_{iy}\omega _e}{\omega _e^2}
\end{equation}

\begin{equation}
u_{iy}=\frac{(f_{ey}+f_{iy})\nu_{ei} - f_{ix}\omega _e}{\omega _e^2} - v_s
\end{equation}

Where the velocity $v_{s}=|\frac{\bf{E}\times \bf{B}}{\bf{B}^2}|$. So we can get,

\begin{equation}
u_{ix}-u_{ex}=\frac{f_{ey}+f_{iy}}{\omega _e}
\end{equation}

\begin{equation}
u_{iy}-u_{ey}=-\frac{f_{ex}+f_{ix}}{\omega _e}
\end{equation}

And if we use equation (41) subtract (42) directly we can get the velocity difference of electrons and ions in the direction $z$:
\begin{equation}
u_{iz}-u_{ez}=\frac{e}{\nu_{ei}m_{i}}E_{z}+\frac{1}{2\nu_{ei}}(f_{iz}-f_{ez})
\end{equation}

By multiplying $ne$, we can get the electric current in the plane perpendicular to the magnetic field.
\begin{equation}
j_{\perp}=j_x  =  ne \frac{f_{ey}+f_{iy}}{\omega _e}
\end{equation}

\begin {equation}
j_{H}=j_y  = -ne\frac{f_{ex}+f_{ix}}{\omega _e}
\end{equation}

\begin{equation}
j_{\parallel}=j_{z} = \frac{ne^{2}}{\nu_{ei}m_{e}}E_{z}+\frac{ne}{2\nu_{ei}}(f_{iz}-f_{ez})
\end{equation}

The velocity of the plasma can be decided by the definition:

\begin{equation}
\bf{v} = \frac {m_e\bf {v}_e + m_i \bf {v}_i}{m_e + m_i}.
\end{equation}

The electric current $j_{\perp}$ is in the direction of electric field perpendicular to magnetic
field, or Pedersen current, and $j_{H}$ is the Hall current. The solution suggests that the current
has no relationship with the collisions and electric field in the plane perpendicular to the magnetic field. The electric current depends on the external forces and magnetic field only. When the external forces are ignored, the electric current in the plane perpendicular to magnetic field disappear.

If we keep the pressure gradient as the only external force, then
$\bf{f}_{i}=\frac{-\bigtriangledown P_{i}}{nm_{e}}$, $\bf{f}_{e}=\frac{-\bigtriangledown P_{e}}{nm_{e}}$. Supposing $P=P_{i}+P_{e}$, we can have the electric currents as:
\begin{equation}
j_{\perp}=j_{x}=- \frac{e(\bigtriangledown P)_{y}}{m_e\omega _e}
\end{equation}
\begin{equation}
j_{H}=j_{y}=\frac{e(\bigtriangledown P)_{x}}{m_e\omega_{e}}
\end{equation}

The above equations show the current density in the whole space, where the plasmas are fully ionized and the gas gradients are considered.

\section{The Energy transport in plasma}

From the discussions in section 3 and section 4, we can find that the electric field perpendicular to the magnetic field cannot cause macroscopic electrical current in plasma. This lead to the failure of the concept of magnetic frozen-in and so do the dynamo and magnetic reconnection, which are widely applied in space science and astrophysics.

One may worry about the fact that the energy released and eruption concerning the magnetic field, such as the plasma heating and the magnetic activities in solar physics and space science. However, we can easily understand the physics processes in plasma with out the concept of ``magnetic frozen-in''.

Recalling the charged particle drift in electromagnetic field, we notice that the drift velocities of the particles represent the motion of the guiding center. The drift velocity depends only on the electric field and magnetic field, and is not related to the initial velocities of the particles. Eventually, the particle is experiencing the motion of drift and Lamor motion. The Lamor velocity depends on the initial velocity of the particle. For the collective effect of the particles, the Lamor motion inside the particles will increase the thermal velocities of the particles through collision, it does not affect the macroscopic motion.

Now we consider the plasma with initial velocity enter the region with magnetic field. The external electric field is always screened. The plasma will be captured by the magnetic field, the drift velocity of the plasma is zero perpendicular to the magnetic field. The kinetic energy of the plasma perpendicular to magnetic field is transformed to the thermal energy of the plasma. With $\triangle\varepsilon$ representing the increase of the thermal energy of the plasma in unit volume, we can get:

    \begin{equation}
    \triangle\varepsilon =  \frac 12 \rho v_{\bot}^2
    \end{equation}

    where $v_{\bot}$ is the velocity of plasma perpendicular to magnetic field, $\rho$ is the density of the plasma.

    Because there is no electric current generated in the processes, the external magnetic field does not change by the motion of plasma.

When plasma is in varied magnetic field, an induced electrical field will be in the plasma. From Farady's law, $\bigtriangledown \times {\bf E} = \frac {\partial {\bf B}}{\partial t}$, the induced electric field will be always perpendicular to the magnetic field. In basic electromagnetism, the variation of the magnetic field will cause eddy electric field, ``the eddy current dissipation'' will happen in the plasma. In metal conductor, the eddy current dissipation is owing to the small resistivity and strong eddy current.

In fact, the eddy current is not necessary for the dissipation. The induced electric field directly contribute energy to the thermal energy of the conductor or plasma in the varied magnetic field. Generally, the charge particles in electric field will gain kinetic energy and form electric current. However, the existence of the magnetic field will act on the current and turn the current into small circles. The average effect of the current will be zero.  But the kinetic energy of the particles will be transformed into the thermal energy due to the collision of the particles. The final effect of the induced electric field for the particles is heating the plasma or conductor.

 In plasma, the induced electric field is always perpendicular to magnetic field, so the current by the induced electric field will be turn around by the magnetic field and forms small circles. The radius of the current is much small. There will be no macroscopic current due to the overlap of the circled electric current.

However, the induced electric field will contribute energy to the thermal energy of the plasma. Since the induced electric field depends on the variation of magnetic field, the increase of the thermal energy of the plasma will come from the variation of the magnetic field. We can suppose the increase of the thermal energy as:
\begin{equation}
\triangle\varepsilon
\sim \frac {\partial B}{\partial t}
\end{equation}
 
Though the transfer of magnetic field to thermal energy may depends on the collisions between charged particles, there is no macroscopic electric current inside the plasma.

From the above discussions, we can conclude that the heating of the plasma in magnetic field may comes from the motion of plasma across magnetic field, and the variation of magnetic field.

Eventually, the processes of plasma heating by variation of magnetic field is observed in the solar atmosphere.  \citet{b18} studied the relation of CME with magnetic field variation. Then found that all of the CME have the variation of magnetic field and maybe the source of CME eruption.

\section{Conclusions}

    In the present paper, we get the electric current and the plasma velocity in the fully ionized plasma by solving the dynamic equations of charged particles. The macroscopic electric current is decided by the difference of the averaged velocities of ions and electrons.   The electric current and plasma velocity is completely decided by electromagnetic field and external forces.

    When the external forces $f_{i}$ and $f_{e}$ acting on ions and electrons respectively,  including the pressure
gradient, the friction and gravitation, are considered, the electric current in the plane perpendicular
to magnetic field will be the form of equation (50) to (52), the electric field perpendicular to magnetic field does not contribute to the macroscopic electric current .  The macroscopic electric current depends only on the external forces.

   The electric current perpendicular to the magnetic field do not depend on the electric field. The induced electric field in magnetic field do not contribute to the electric current, since the induced electric field is always perpendicular to the magnetic field. In any reference system with velocity $\bf{v}'$, the induced electric field $\bf{v}'\times \bf{B}$ is perpendicular to the magnetic field, and will not contribute to the macroscopic electric current, so cannot distort the magnetic field in space. The velocity of the plasma is not coupled with magnetic field.

When the external forces are ignored, there is no electric current in the plane perpendicular to magnetic field. The Pedersen current and Hall current are zero. The plasma has a velocity,
\begin {equation}
\bf{u}=\frac{\bf{E}\times \bf{B}}{B^{2}}
\end {equation}
It is the global drift of the plasma in electromagnetic field. In plasma, the electric field is screened, the drift velocity is zero.

When plasma with average velocity $v$ enter magnetic field, the plasma will be confined by the magnetic field. The kinetic energy of the plasma perpendicular to magnetic field will be transported into plasma thermal energy. The macroscopic velocity perpendicular to magnetic field is zero. It is one process for plasma heating.

When plasma is in varied magnetic field, the induced electric field will heating plasma. The macroscopic electric current is not presented for the existence of magnetic field. The heating process is not from Ohm's dissipation, but directly from the action of induced electric field and magnetic field. The increase of the plasma thermal energy depends on the variation of magnetic field with time.

In this paper, we don't consider the boundary condition of the plasma and the inhomogeneous magnetic field. The effect of the boundary and inhomogeneous field will have effect on the electric current in the plasma. The work will be done in future research.


\begin{thebibliography}{99}
\bibitem[\protect\citeauthoryear{Falthammar}{2006}]{b01} Falthammar, C-G., 2006, Comments on the motion of magnetic field lines,  Am. J. Phys. 74(5), 454-455

\bibitem[\protect\citeauthoryear{Alfven}{1976}]{b02} Alfven, H., 1976, On frozen-in field lines and field-line reconnection, J. Geophys. Res. 81, 4019-4021

\bibitem[\protect\citeauthoryear{Alfven}{1942}]{b1} Alfven, H.,1942, Existence of electromagnetic-hydrodynamic waves, Nature, vol. 150,405

\bibitem[\protect\citeauthoryear{Lighthill}{1960}]{b2} Lighthill, M.J. 1960, Studies on magneto-hydrodynamic waves and other anisotropic wave motions, Phil. Trans. Roy. Soc.
London, Vol. 252A, pp.397-430

\bibitem[\protect\citeauthoryear{Witalis}{1986}]{b3}  Witalis, E.A., 1986, Hall magnetohydrodynamics and its applications to laboratory and cosmic plasma, IEEE Transactions
on plasma science, Vol.PS-14, No.v 6, 842-848

\bibitem[\protect\citeauthoryear{Spitzer}{1962}]{b4} Spitzer, L., Jr. 1962, Physics of fully ionized gases, second revised edition, Dover Publications, Inc. Mineola, New York

\bibitem[\protect\citeauthoryear{Somov}{2000}]{b5} Somov, B. V., 2000, Cosmic Plasma Physics, Kluwer Academic Publishers, Dordrecht, p 181-190

 \bibitem[\protect\citeauthoryear{Volkov}{1966}]{b6} Volkov, TF., 1966, Hydrodynamic description of a collisionless plasma, in Reviews of plasma physics, ed. M.A. Leontovich, New
York, Consultant Bureau, v. 4, 1-21

\bibitem[\protect\citeauthoryear{Sweet}{1958}]{b7} Sweet, P.A., 1958, the neutral point theory of solar flares, in :Electromagnetic phenomena in Cosmical Physics, edited by:
Lehnert, B., Cambridge University Press, London, 1958

\bibitem[\protect\citeauthoryear{Parker}{1957}]{b8} Parker, E. N., 1957, Sweet's mechanism for merging magnetic fields in conducting fluids, J. Geophys. Res., 62, 509-520

\bibitem[\protect\citeauthoryear{Dungey}{1961}]{b9}  Dungey, J. W., 1961, Interplanetary fields and the auroral zone, Phys. Rev. Lett. 6, 47-48

\bibitem[\protect\citeauthoryear{Paschmann et al}{1979}]{b10}  Paschmann, G, O. sonnerup, B. U., Papamastroakis, I., Sckopke, N., Haerendel, G., Bame, S.J., Asbridge, J. R., Gosling, J.T.,
Russell, C.T., and Elphic, R.C., 1979, Plasma acceleration at the earth's magnetopause: Evidence for reconnection,
nature Lond. 282, 243-246

\bibitem[\protect\citeauthoryear{Paschmann \& Treumann}{2003}]{b11} Paschmann, G. and Treumann, R., 2003, Auroral Plasma Physics, edited by: Paschmann, G., Haaland, S., Treumann, R.,
ISSI Space Science Series, Vol 15, ISBN 1-4020-0936-1

\bibitem[\protect\citeauthoryear{Yamauchi \& Blomberg}{1997}]{b12} Yamauchi, M. and Blomberg, L., 1997, Problems on mappings of the convection and on the fluid concept, Phys. Chem. Earth. 22,
709-714

\bibitem[\protect\citeauthoryear{Parker}{1979}]{b13} Parker, E. N., 1979, Cosmical magnetic fields, Clarendon Press, Oxford, p31-32

\bibitem[\protect\citeauthoryear{Gomobsi}{1994}]{b14} Gomobsi, T. I., 1994, Gaskinetic Theory, Cabridge Univ., New York

\bibitem[\protect\citeauthoryear{Song, Gombosi \& Ridley}{2001}]{b15}  Song, P., Gombosi, T. I., Ridley, A. J., 2001, Three-fluid Ohm's law, Journal of Geophysical Research, Volume 106, Issue
A5, p. 8149-8156

\bibitem[\protect\citeauthoryear{Pandey \& Wardle}{2008}]{b16} Pandey, B.P. and Wardle, M., 2008, Hall magnetohydrodynamics of partially ionized plasmas, Mon. Not. R. Astron. Soc., 385,
2269-2278

\bibitem[\protect\citeauthoryear{Chen}{1974}]{b17}  Chen, F.F., 1974, Introduction to plasma physics, Plenum Press

\bibitem[\protect\citeauthoryear{Zhang, Zhang \& Zhang}{2007}]{b18} Zhang, Y., Zhang, M. \& Zhang, H. Q., 2007, A statistical study on the relationship between surface field variation and CME initiation, Advances in Space Research, Vol. 39, P1762-1766

\end{thebibliography}
\end{document}